\documentclass[12pt]{article}
\usepackage[english]{babel}
\usepackage{hyperref,hypcap,ae,tikz}
\usepackage{graphicx}
\usepackage{bm}
\usepackage{amsmath}
\usepackage{mathrsfs}
\usepackage{mathtools}
\usepackage{amsfonts}
\usepackage[T1]{fontenc}
\usepackage{textcomp}
\usepackage{xcolor}
\usepackage{lmodern}
\usepackage[caption=false]{subfig}
\usepackage{booktabs}
\usepackage{cite}
\usepackage[section]{placeins}
\usepackage{authblk}
\hypersetup{
	colorlinks=true,
	citecolor=blue,
    linkcolor=magenta,
    urlcolor=blue,
	}

\begin{document}

\title{{\bf Arrow of time problem in gravitational collapse}}

\author[1]{Samarjit Chakraborty \thanks{\href{mailto:samarjitxxx@gmail.com}{samarjitxxx@gmail.com}}}
 \author[1]{Sunil D. Maharaj \thanks{\href{mailto:MAHARAJ@ukzn.ac.za}{MAHARAJ@ukzn.ac.za}}}
\author[1]{Rituparno Goswami \thanks{\href{mailto:Goswami@ukzn.ac.za}{Goswami@ukzn.ac.za}}}
\author[2]{Sarbari Guha \thanks{\href{mailto:srayguha@yahoo.com}{srayguha@yahoo.com}}}
\affil[1]{Astrophysics Research Centre, School of Mathematics, Statistics and Computer Science, University of KwaZulu-Natal, Private Bag X54001, Durban 4000, South Africa.}
\affil[2]{Department of Physics, St.Xavier's College (Autonomous), Kolkata 700016, India.}

\date{}

\maketitle

\section*{Abstract}
We investigate the arrow of time problem in the context of gravitational collapse of radiating stars in higher dimensions for both neutral and charged matter. The interior spacetime is described by a shear-free spherically symmetric metric filled with a dissipative fluid. The exterior spacetime of the radiating star is taken as the higher dimensional Vaidya metric. We establish that the arrow of time associated with gravitational entropy is opposite to the thermodynamic arrow of time for all dimensions. The physical consequences of our results are considered. Our result conforms with previous studies on shear-free spherical collapse, which suggests, avoidance of the naked singularity as the end state results in a wrong arrow of time, indicating a fundamental problem with the local application of the Weyl curvature hypothesis.

\vspace{0.1cm}
Keywords: Gravitational entropy, Weyl curvature hypothesis, arrow of time, radiating stars, higher dimensions.

\section{Introduction}
 This paper deals with the arrow of time problem in a shear-free radiating gravitational collapse in $ N $ dimensions. In the subsequent portions of this section we introduce the problem with proper context and motivation.
\subsection{The context: Weyl curvature hypothesis}
The entropy of the universe should increase with time according to the second law of thermodynamics (SLT). However, the cosmic microwave background radiation (CMBR) indicates a state of thermal equilibrium during the initial phase of the universe. This suggests that the universe began evolving from a maximum state of entropy, which clearly violates the SLT. To resolve this issue Penrose \cite{Penrose1,Penrose2} proposed the idea of Weyl curvature hypothesis (WCH). According to the WCH most of the entropy of the universe is carried by the free gravitational field, which is called gravitational entropy (GE) (associated with the degrees of freedom of the free gravitational field) and it is a function of the Weyl curvature tensor. Therefore, the total entropy of the universe (which is the sum of the conventional thermodynamic entropy and the gravitational entropy) was very low at the initial phase of the cosmological evolution, and with the formation of structures in the universe, the gravitational entropy increased monotonically \cite{Bolejko} giving rise to clumping of matter and inhomogeneity (due to gravitational tidal forces).

\subsubsection{Arrow of time problem: global vs local}
Consequently,  such an evolution of structures in the universe produces the observed arrow of time associated with the gravitational entropy. Largely, on a cosmological scale, the thermodynamic arrow of time associated with radiation coincides with the gravitational arrow of time. Locally the gravitational entropy should reproduce the Hawking-Bekenstein entropy for black holes (BH), making BH entropy a special case of gravitational entropy. This makes sense because most of the entropy of the universe is BH entropy, which is area dependent (unlike matter entropy) \cite{Penrose2}. Similarly, one can think that most of the entropy in the universe is gravitational entropy.

However, the local arrow of time can be understood if we study the process of collapse, i.e., before the formation of BH or other astrophysical objects. It is reasonable to think that during the process of stellar collapse the interior gravitational arrow of time should match the outside radiation arrow. However, Bonnor in \cite{Bonnor1,Bonnor2,Bonnor3} found that this is not so. The gravitational arrow of time is oriented opposite to that of the outside radiation. This is a brief explanation of the so-called "Arrow of time problem" in collapse.

\subsubsection{ Gravitational epoch functions}
The WCH needs some quantitative measure, these are called gravitational epoch functions. In other words, $\mathcal{P}(x^i)$ is called a
gravitational epoch function of an event $x^i$ and if it is to obey the
Penrose’s hypothesis, $ \mathcal{P} $ should be a non-decreasing function of time, i.e., $\partial \mathcal{P}/\partial t \geq 0$. Therefore, one can also make a measure of GE using $ \mathcal{P} $. Essentially the $ \mathcal{P} $ function measures the relative strength of the Weyl curvature (due to free gravity field through the Weyl tensor $ C_{abcd} $) to the curvature due to matter (via the Ricci tensor $ R_{ab} $). 

Following the Weyl curvature hypothesis many different kinds of gravitational epoch functions were proposed to measure GE. One such epoch function was used by Bonnor \cite{Bonnor1,Bonnor2,Bonnor3} in a four dimensional dissipative collapse setting, where he showed that the local gravitational arrow of time is opposite to the thermodynamic (radiation) arrow of time. Bonnor used the gravitational epoch function $ \mathcal{P} $, defined as
 \begin{equation}\label{P}
\mathcal{P}={(C_{abcd}C^{abcd})}{(R_{ab}R^{ab})}^{-1}={\mathcal{W}}/{\mathcal{R}},
\end{equation}
where  $ \mathcal{W}\equiv C_{abcd}C^{abcd}$, denotes the square of the Weyl curvature tensor and $\mathcal{R}\equiv R_{ab}R^{ab}$ is the Ricci tensor square. In a recent study \cite{Arrow1} we have checked the validity of Bonnor's findings, and extended this claim using the gravitational entropy epoch function, proposed by Rudjord et al \cite{entropy1,entropy2}. In this case the gravitational epoch function $\mathcal{P}_{1}$ is given by
\begin{equation} \label{P1}
\mathcal{P}_{1}={(C_{abcd}C^{abcd})}{(R_{abcd}R^{abcd})}^{-1}=\mathcal{W}/\mathcal{K},
\end{equation}
where $R_{abcd} $ represents the Riemann tensor, which contains both the trace part (Ricci tensor) and the traceless part (Weyl tensor) with $\mathcal{K}\equiv R_{abcd}R^{abcd} $ as the Kretschmann scalar.

According to the WCH, the Ricci part dominates over the Weyl part during the initial phase of the universe, and with evolution the Weyl part takes over the Ricci part, giving rise to the arrow of time in the universe due to increasing gravitational entropy. However, in a collapse situation it fails to give the correct arrow of time. This raises questions on the applicability of the Weyl curvature hypothesis on a local scale.

It is important to mention some features of the above mentioned gravitational epoch functions. The function $\mathcal{P}$ is well defined in non-vacuum models but fails in empty space because the Ricci tensor vanishes. For example, in external Schwarzschild spacetime, the Ricci tensor is zero, and  $ \mathcal{P} $ is undefined. This is where $ \mathcal{P}_{1} $ becomes useful and is well defined in such situations. In addition to the information contained in $\mathcal{P}$, the epoch function $\mathcal{P}_{1}$ contains the contribution from the Ricci scalar also, which indicates how the spacetime differs from Euclidean space. Both $\mathcal{P}$ and $\mathcal{P}_{1}$ vanish in the Friedmann-Robertson-Walker models (and for all conformally flat spacetimes) of standard cosmology, as the Weyl tensor vanishes in these models.

 Using $\mathcal{P}_1$, we can compute the gravitational entropy of a black hole \cite{entropy1} and match it with the Hawking-Bekenstein entropy \cite{Bekenstein,SWH1,Chandra}.  This helped Rudjord et al \cite{entropy1,entropy2} to fix the constants in the GE and to calculate the entropy of Schwarzschild black hole and the Schwarzschild-de-Sitter spacetime. There have been many studies \cite{gc,cgg,cgg1} where the gravitational entropy is determined for local astrophysical objects, and they are shown to be valid using the Weyl proposal due to Rudjord et al \cite{entropy1,entropy2} and the Clifton-Ellis-Tavakol (CET) proposal \cite{CET}.
In a non-vacuum $N$ dimensional spacetime the two proposals differ from each other in the following manner
\begin{equation}
\mathcal{P}_{1}^{-1}-\frac{4}{N-2}\mathcal{P}^{-1}=1-\frac{2}{(N-1)(N-2)}\frac{R^2}{\mathcal{W}} ,
\end{equation}
where $ R $ is the Ricci scalar. This is true for any dimension $ N\geq4 $.

\subsubsection{Key question: Is the arrow of time in collapse dimension dependent?}

It is known that the curvature scalars are dimension sensitive \cite{misha} and therefore, there is a good possibility that the behaviour of gravitational entropy  will differ in higher dimensions. The number of spacetime dimensions plays a crucial role in higher dimensional general relativity, modified gravity theories, string theories and other theories of gravity. Several authors have studied the Schwarzschild, Reissner-Nordström, de Sitter/anti de Sitter and Kerr spacetimes in higher dimensions\cite{[1],[2],[3],[4],[5]}. The studies show that the higher dimensional de Sitter space is a Lorentzian analogue of a $N$-sphere. In \cite{[6]} it is shown that the charge reduces the range of stable parameters and the damping rate of stable perturbations. Also, the $N$ dimensional Kerr-Newman spacetime is still not well understood. 

The existence of higher dimensions also influences the dynamics of compact astrophysical objects. It has been observed that the mass-radius ratio of compact objects, and also their stability depends on the number of dimensions. Several authors have found bounds on the mass-radius ratio and its dimensional dependence in general relativity \cite{[7],[8],[9],[10]}. 

In this study we are interested in radiating stars and several models have been found in four dimensions. Santos \cite{[11]} deduced the boundary conditions at the stellar surface that must be satisfied for such a radiating system. Subsequently, Maharaj et al. extended this condition for a generalised exterior and an interior with a composite matter distribution \cite{[12],[13]}. This composite matter distribution can also carry charge \cite{[14]} and modifies the pressure at the stellar boundary, generalising earlier works \cite{[15],[16],[17]}. It is clear that the number of spacetime dimensions are crucial in modelling higher dimensional radiating stars.

 Consequently, researchers have tried to find a higher dimensional counterpart of the Santos condition, as in \cite{[18],[19],[20]}, where an uncharged shear-free interior is matched with the conventional pure Vaidya exterior spacetime. However, we need the higher dimensional Vaidya metric as the exterior to fully understand the higher dimensional radiating collapse. This spacetime geometry is very important in the field of relativistic astrophysics, especially for studying the end state of highly dense objects, gravitational collapse and the formation of singularities \cite{[21],[22],[23],[24],[25],[26],[27],[28],[29],[30],[31],[32],[48]}. Thermodynamical studies of such radiating fluids have also been studied by Debnath \cite{[33]}. 
 
 Researchers have also used higher dimensional modified theories of gravity for this study. For example in \cite{[34],[35],[36],[37]}, using the generalised Vaidya analogue spacetime in EGB gravity, gravitational collapse have been studied. Other researchers have also investigated collapse using extensions of the generalised Vaidya spacetime \cite{[38],[39],[40]}. The most general study, which contains all the previous results, was done by Maharaj et al. \cite{byron}, where a charged composite interior is matched to the higher dimensional generalised Vaidya spacetime at the stellar surface in higher dimensions and deduced the most general matching conditions in $N$ dimensions in general relativity. Therefore, it was found that, in a radiating collapse of a star the number of spacetime dimensions directly affect the matching conditions at the boundary surface.

In this scheme of things the investigation of the arrow of time problem in $  N$ dimensional gravitational collapse of a radiating star becomes extremely significant. We mention some of the important works on GE. A detailed study on GE for Lema\^{i}tre-Tolman-Bondi (LTB) dust collapse, cosmic voids and other systems was conducted by Sussman and others in \cite{suss1,suss2,suss3,HB,LNP,Piza}.

\subsection{This paper}
In this study we consider the collapse of a shear-free radiating star. We extend our previous study on gravitational entropy in collapse \cite{Arrow1} for higher dimensions. The main idea is to examine the orientation of the gravitational arrow of time in higher dimensions with respect to the well established radiation (thermodynamic) arrow. Because the radiation emanating from the star always moves outwards, it provides us a clear sense of the thermodynamic arrow of time. The study of such a higher dimensional system is very important to check whether the gravitational arrow of time changes direction depending upon the spacetime dimensions.

This paper is organised as follows:
In the following section we briefly present the model of a radiating collapsing star in $N$ dimensional general relativity by taking the interior spacetime as a shear-free spherically symmetric metric and the exterior is taken as the generalised Vaidya metric. Two shear-free spherically symmetric interior spacetimes are explicitly studied for the arrow of time problem. The first spacetime has time independent lapse function and the second one has a time dependent lapse function. For both the cases we present our conclusions in detail. We have also studied the first spacetime when charge is present. Finally we conclude with a discussion on various aspects of our findings.

\section{Radiating collapse in $N$ dimensions: overview}
For a collapsing radiating star we can describe the system in two parts \cite{[11],byron,olivera1,olivera2}: the interior metric representing a collapsing sphere of fluid with radial heat flow and the exterior metric given by the Vaidya metric with outward radiation. This provides us with a sense of the thermodynamic arrow of time.

The interior of this system can be described by the Einstein field equations with a dissipative fluid
\begin{equation}\label{ein}
\begin{aligned}
 R_{ab}-\frac{1}{2}\textsl{g}_{ab}R =G_{ab}=\kappa_{N}T_{ab},\,\,\,\,\,\,\,\,\,\,\,\,\,\,\,\,\,\\ 
 T_{ab} =[(\mu + p)v_{a}v_{b}+p \textsl{g}_{ab}+q_{a}v_{b}+q_{b}v_{a}+\pi_{ab}],
\end{aligned}
\end{equation}
where $\kappa_{N}$ is the coupling constant \cite{[42],[43]} in higher dimensions $ N\geq 4 $ given by the following expression
\begin{equation}
\kappa_{N}=\dfrac{2(N-2)\pi^{\frac{N-1}{2}}}{(N-3)\left(\frac{N-1}{2}-1\right)!}.
\end{equation}
This represents a fluid with energy density $ \mu $, pressure $ p $, heat flow $ q^{a}=(0,q,0,.\,.\,.\,,0) $, four-velocity $ v_{a} $ and anisotropic stress tensor $ \pi_{ab} $. The interior spacetime of the star is represented by a spherically symmetric metric. The general spacetime is given by
\begin{equation}\label{int}
ds_{-}^2=-A(r,t)^2 dt^2+B(r,t)^2 dr^2+Y(r,t)^2d\Omega_{N-2}^{2},
\end{equation}
where
\begin{equation}
d\Omega_{N-2}^{2}=\sum_{i=1}^{N-2} \left[\prod_{j=1}^{i-1} \sin^2(\theta_{j})\right](d\theta_{i})^{2}.
\end{equation}
Therefore, we can take the four-velocity of the fluid in comoving coordinates as $v^{a}=(1/A,0,0,.\,.\,.\,,0)$. Consequently, here the kinematical quantities in $N$ dimensions become 
\begin{equation}
\begin{aligned}
a^{a}&=\left(0,\frac{A^{'}}{AB^{2}},0,0,.\,.\,.\,,0\right),\,\,\, \Theta =\frac{1}{A}\left(\frac{\dot{B}}{B}+(N-2)\frac{\dot{Y}}{Y}\right),\\
\sigma &= \frac{1}{\sqrt{N-1}A}\left(\frac{\dot{Y}}{Y}-\frac{\dot{B}}{B}\right),\, \omega_{ab} =0,
\end{aligned}
\end{equation}
where $ a^{a} $ is the fluid $ N $-acceleration, $ \Theta $ is the expansion scalar, $ \sigma $ is the shear and $ \omega_{ab} $ is the vorticity tensor. In the above expressions, dots represent time derivatives and primes denote radial derivatives. All the kinematical quantities are related via the following relation in $ 1+3 $ decomposition:
\begin{equation}
\nabla_{b}u_{a}=-u_{a}a_{b}+\frac{1}{N-1}h_{ab}\Theta+\sigma_{ab}+\omega_{ab},
\end{equation}
where $\nabla $ is the covariant derivative and $ h_{ab}=\textsl{g}_{ab}+u_{a}u_{b} $ is the projection tensor. The anisotropic stress tensor is related to the difference of radial pressure $p_{r}$ and tangential pressure $ p_{t} $ as 
\begin{equation}
\pi_{ab}=\Delta(\mathcal{N}_{a}\mathcal{N}_{b}-(N-1)^{-1}h_{ab}),
\end{equation}
where $ \Delta=p_{r}-p_{t} $. Consequently, the isotropic pressure can be expressed as
\begin{equation}
p=(N-1)^{-1}\left(p_{r}+(N-2)p_{t}\right)
\end{equation}   
in $ N $ dimensions. Here $\mathcal{N}_{a}  $ is a spacelike unit vector of the form $\mathcal{N}_{a} =(0,B^{-1},0,0,.\,.\,.\,,0)$. In our study we will only consider isotropic pressure, i.e., $ p_{r}=p_{t}=p $ for our calculations.

The static interior metric with $A_{0},B_{0}$ is like a Schwarzschild interior \cite{ray} with a perfect fluid (without heat flow) which is to match the Schwarzschild exterior spacetime (to which \eqref{ext} reduces if the mass $ m $ is constant). However, with \eqref{int_soln} the solutions become radiative. To match spacetimes with heat flow we need a time-dependent extension of the Schwarzschild exterior metric. For exact functional forms of $A_{0}$ and $ B_{0} $ please see \cite{gold1,gold2}.

To ensure the physicality of the collapsing matter field, we consider it to obey the usual weak energy condition (WEC), i.e., the energy density measured by any timelike observer is non-negative. We can also consider the dominant energy condition (DEC), i.e., the local energy flow is nonspacelike to a timelike observer. These two energy conditions ensure that the collapsing matter fluid remains physically reasonable. Therefore $ T_{ab}v^{a}v^{b} $ has to be non-negative and $ T^{ab}v_{b} $ is nonspacelike for any timelike vector $ v^{a} $.  In brief the energy conditions give us the following
\begin{align}
\mathcal{D}> 0,\,\,\mu \geq p,\,\,\,\mu+\mathcal{D}\geq 3p,\,\,\, \mu +p+\mathcal{D}\geq 0,\,\, \mathcal{D}^2\equiv(\mu +p)^2 -4(qB)^2.
\end{align}
We can also take $ \mu \geq 0 $ for a physically reasonable matter field. For more details please see \cite{engycon}.

The exterior metric represents an outgoing radiation field around a spherically symmetric star which can be assumed to be the generalised Vaidya metric in $N$ dimensions described by
\begin{equation}\label{ext}
ds_{+}^2=-\left(1-\frac{2m(u,R)}{(N-3)R^{N-3}}\right)du^2-2dudR+ R^2d\Omega_{N-2}^{2},
\end{equation}
where $m(u,R)$ is the higher dimensional Misner-Sharp mass containing the gravitational energy within the radius $ R $. The four-dimensional mass function is given in \cite{[49],[50]}.

We consider a timelike $ N-1 $ dimensional spherical hypersurface $\Sigma $, which is the surface of the star. We can match both the interior and exterior  spacetimes using the two junction conditions and finally get the following relation on $ \Sigma $:
\begin{equation}\label{jn1}
p_{\Sigma}=(qB)_{\Sigma}=(qB_{0}f)_{\Sigma}.
\end{equation}
Therefore the pressure at the boundary is not zero for a radiating star. For a detailed analysis of matching conditions see \cite{byron,[11],[44],[45],[46],[47]}.

\section{Arrow of time problem in spherically symmetric shear-free spacetime collapse}

In this section we consider two spherically symmetric  shear-free spacetimes and study the arrow of time problem in a collapsing scenario.

\subsection{The first spacetime: Time-independent lapse function}

We consider the shear-free metric for the interior spacetime, which starts from a static solution and dynamically evolves with time. Consequently, we adopt the following conditions 
\begin{equation}\label{int_soln}
A=A_{0}(r),\;\;Y=rB,\;\; B=B_{0}(r)f(t);\;\; f(t)>0,\;\; A_{0}>0,\;\; B_{0}>0 ,
\end{equation}
where $ A_{0}, \, B_{0} $ are solutions of the interior metric for the static perfect fluid without heat flow, having $ \mu_{0} $ as the energy density and $ p_{0} $ as the isotropic pressure. Further, $ f(t) $ is a positive function to be determined \cite{Bonnor3}. A detailed study on heat conducting shear-free fluids was completed by Sussman in \cite{suss0}. For recent treatments of the geometrical properties of the metric functions \eqref{int_soln} in the context of the radiating stars see Paliathanasis et al \cite{X1} and Ivanov \cite{X2}.

Then the field equations for the interior spacetime \eqref{int_soln} become
\begin{align}
\kappa_{N} \mu A_{0}^2=&\; \frac{\dot{f}^2}{f^2}\left[\frac{(N-1)(N-2)}{2}\right]
 \nonumber\\
& -\frac{A_{0}^2}{f^2 B_{0}^2}\left[(N-2)\frac{B_{0}^{''}}{B_{0}}+\frac{(N-2)^{2}B_{0}^{'}}{rB_{0}}+\frac{(N-2)(N-5)}{2}\frac{{B_{0}^{'}}^2}{B_{0}^2}\right],  \nonumber\\
-\kappa_{N} q B_{0}^2=&\;(N-2)\frac{A_{0}^{'}}{A_{0}^2}\left(\frac{\dot{f}}{f^3}\right), \nonumber
\end{align}
\begin{equation}\label{shfrein}
\begin{aligned}
\kappa_{N} p =&-\frac{1}{A_{0}^2}\left[(N-2)\frac{\ddot{f}}{f}+\dfrac{(N-2)(N-3)}{2}\left(\frac{\dot{f}}{f}\right)^2\right] \\
& +\dfrac{1}{B_{0}^{2}f^{2}} \left[ \dfrac{(N-2)(N-3)}{2}\Biggl\{\left(\frac{B_{0}^{'}}{B_{0}}\right)^2+ \frac{2}{r}\frac{B_{0}^{'}}{B_{0}}\Biggl\}+(N-2)\Bigg(\frac{A_{0}^{'}}{A_{0}}\frac{B_{0}^{'}}{B_{0}}+\frac{1}{r}\frac{A_{0}^{'}}{A_{0}} \Bigg) \right], \\
\kappa_{N} p B_{0}^2 =&-\frac{B_{0}^{2}}{A_{0}^{2}}\left[(N-2)\frac{\ddot{f}}{f}+(N-3)\left(\frac{\dot{f}}{f}\right)^{2}\right] \\
& +\dfrac{1}{f^2}\left[\frac{A_{0}^{''}}{A_{0}}+(N-4)\frac{A_{0}^{'}B_{0}^{'}}{A_{0}B_{0}}+(N-3)\Biggl\{\frac{1}{r}\left(\frac{A_{0}^{'}}{A_{0}}+\frac{B_{0}^{'}}{B_{0}}\right)-\frac{{B_{0}^{'}}^2}{B_{0}^{2}}+\frac{B_{0}^{''}}{B_{0}}\Biggl\}\right] \\
& -\frac{(N-3)(N-4)}{2f^2}\left[\frac{B_{0}^{2}\dot{f}^{2}}{A_{0}^2}-\frac{B_{0}^{'}}{B_{0}}\left(\frac{2}{r}+\frac{B_{0}^{'}}{B_{0}}\right)\right].
\end{aligned}
\end{equation}
From \eqref{shfrein}, it follows that the isotropic pressure $ p $, energy density $ \mu $ and heat flow $ q_{a} $ for the interior spacetime \eqref{int}, with the conditions \eqref{int_soln}, are given by the equations
\begin{equation}\label{pr}
p=\frac{p_{0}}{f^2}-\frac{1}{\kappa_{N}A_{0}^2}\bigg\lbrace(N-2)\frac{\ddot{f}}{f}+\dfrac{(N-2)(N-3)}{2}\left(\frac{\dot{f}}{f}\right)^2\bigg\rbrace ,
\end{equation}
\begin{equation}\label{mu}
\mu=\frac{\mu_{0}}{f^2}+\frac{(N-1)(N-2)}{2\kappa_{N} A_{0}^2}\left(\frac{\dot{f}}{f}\right)^2 ,
\end{equation}
\begin{equation}\label{qa}
q^{\alpha}=q\delta_{1}^{\alpha}=-\dfrac{(N-2)A_{0}^{'}}{\kappa_{N} A_{0}^{2}B_{0}^{2}}\left(\frac{\dot{f}}{f^{3}}\right)\delta_{1}^{\alpha} ,
\end{equation}
where the energy density and isotropic pressure for the static solution (without heat flow) are given by $ \mu_{0} $ and $ p_{0} $ respectively. There is no heat flow for the static solution.
Consequently, imposing the condition of vanishing pressure (static solution) $p_{0}\vert_{\Sigma}=0$ on the boundary surface $\Sigma$, and using equations \eqref{jn1}, \eqref{pr}, and \eqref{qa}, we obtain the following second order differential equation for $f(t)$ in $N$ dimensions as
\begin{equation}\label{fnd}
2f\ddot{f}+(N-3)\dot{f}^2-2a\dot{f}=0, \;\;\;\; a=\left(\frac{A_{0}^{'}}{B_{0}}\right)_{\Sigma} ,
\end{equation}
where $ a $ must be positive if the static solution generated by $ A_{0} $ and $ B_{0} $ is to match with the Schwarzschild exterior metric. Subsequently, we can find the first integral of \eqref{fnd} as
\begin{equation}\label{dotfN}
\dot{f}=\frac{1}{(N-3)}\left(2a-C_{1}f^{-\frac{(N-3)}{2}}\right),
\end{equation}
which integrates to give the quadrature
\begin{equation}
t=\frac{1}{a}\left(\frac{1}{2a}\right)^{\frac{2}{N-3}}\int\frac{(u+C_{1})^{\frac{2}{N-3}}}{u}du+C_{2}.
\end{equation}

We have tabulated all the solutions of the equation \eqref{fnd} in Table \ref{tabf} for various higher dimensions. We observe that as the dimension $N$ increases the analytical solution of the second integral becomes very difficult. Here $ C $ and $ C_{2} $ are constants of integration with $ \tilde{C}_{N}=C/(N-3) $.
\vspace{1cm}
{\renewcommand{\arraystretch}{3.5}
\begin{table}[hbt!]
   \tiny
   \centering
   \rotatebox{0}{
   \begin{minipage}{\textwidth}
   \centering
   \caption{Nature of the differential equation for $ f $ in higher dimensions in GR} 
    \label{tabf}
    \resizebox{\textwidth}{!}{
   \begin{tabular}{lccr}
   \toprule\toprule
  {\normalsize\textbf{$N$}} & {\normalsize\textbf{Condition on $\Sigma$}} & {\normalsize\textbf{First integral}} & {\normalsize\textbf{Second integral} }\\ 
   \midrule
  $4$ & $2f\ddot{f}+\dot{f}^2-2a\dot{f}=0$ & $ \dot{f}=2a-\tilde{C}_{4}f^{-1/2} $ & $ t=\frac{2a^{2}f+2aC\sqrt{f}-\frac{3C^2}{2}+C^{2}\ln\vert2a\sqrt{f}-C\vert}{4a^{3}}+C_{2} $ \\
   $5$ & $2f\ddot{f}+2\dot{f}^2-2a\dot{f}=0$ & $  \dot{f}=a-\tilde{C}_{5}f^{-1} $ & $t=\frac{f}{a}+\frac{C}{2a^2}\ln\vert 2af-C\vert+C_{2}-\frac{C}{2a^2}  $\\
$6$ & $2f\ddot{f}+3\dot{f}^2-2a\dot{f}=0$ & $  \dot{f}=2a/3-\tilde{C}_{6}f^{-3/2} $ & $ \!\begin{aligned}[t]& t=\frac{3f}{2a}+\frac{3C}{2a^2}\int\frac{\sqrt{f}}{2af^{3/2}-C}dw+C_{2};\\ &w=(2a)^{\frac{1}{3}}f^{\frac{N-3}{6}}\end{aligned} $\\   
$7$ & $2f\ddot{f}+4\dot{f}^2-2a\dot{f}=0$ & $  \dot{f}=a/2-\tilde{C}_{7}f^{-2} $ & $t=\frac{2f}{a}+\frac{\sqrt{C}}{2\sqrt{2a}}\ln\left\vert\frac{\sqrt{2a}f-\sqrt{C}}{\sqrt{2a}f+\sqrt{C}}\right\vert +C_{2} $\\  
$8$ & $2f\ddot{f}+5\dot{f}^2-2a\dot{f}=0$ & $  \dot{f}=2a/5-\tilde{C}_{8}f^{-5/2} $ & $ \!\begin{aligned}[t] &t=\frac{5f}{2a}+\frac{5C}{2a^{2}}\int\frac{\sqrt{f}}{(2af^{5/2}-C)}du+C_{2};\\
& u=2af^{5/2}-C \end{aligned} $\\ 
$9$ & $2f\ddot{f}+6\dot{f}^2-2a\dot{f}=0$ & $  \dot{f}=a/3-\tilde{C}_{9}f^{-3} $ & $ \!\begin{aligned}[t] t & =\frac{3f}{a}+\frac{\sqrt{C}}{a(2a)^{\frac{1}{3}}}(\ln\vert (2a)^{\frac{1}{3}}f-\sqrt{C}\vert \\
& +\frac{1}{\omega^2}\ln\vert(2a)^{\frac{1}{3}}f-\sqrt{C}\omega\vert \\
&+\frac{1}{\omega}\ln\vert(2a)^{\frac{1}{3}}f-\sqrt{C}\omega^{2}\vert)+C_{2};\\
&\omega=e^{\frac{2\pi i}{3}} \text{is a complex cube of unity}.
 \end{aligned}$\\ 
$10$ & $2f\ddot{f}+7\dot{f}^2-2a\dot{f}=0$ & $  \dot{f}=2a/7-\tilde{C}_{10}f^{-7/2} $ & $ \!\begin{aligned}[t]& t=\frac{7f}{2a}+\frac{C}{2a^{2}}\int\frac{\sqrt{f}}{2af^{7/2}-C}du+C_{2};\\
& u=2af^{7/2}-C \end{aligned} $\\ 
   \bottomrule
   \end{tabular}
   }
   \end{minipage}}
\end{table}
}

On physical grounds we assume pressure isotropy, i.e., $ \Delta=0 $. The condition of isotropy of pressure implies the following relation for the interior fluid
\begin{eqnarray}\label{isp}
\frac{A_{0}^{''}}{A_{0}}+(N-3)\frac{B_{0}^{''}}{B_{0}}=
-\dfrac{(N-3)(N-4)}{2}\left(\frac{B_{0}^{'}}{B_{0}}\right)\left(\frac{2}{r}+\frac{B_{0}^{'}}{B_{0}}\right)\nonumber\\
+\left(\frac{A_{0}^{'}}{A_{0}}\right)\left(\frac{1}{r}+2\frac{B_{0}^{'}}{B_{0}}\right)+(N-3)\left(\frac{B_{0}^{'}}{B_{0}}\right)\left[\frac{(N-3)}{r}+\frac{B_{0}^{'}}{B_{0}}\frac{N}{2}\right].
\end{eqnarray}
It is interesting to note that all the temporal contributions exactly cancel and we get a time independent relation. This means that the pressure isotropy is completely determined by the radial functions of the interior metric and the dimension; moreover, the relation holds for all time. In the above equation \eqref{isp} we can see that the first term in the RHS is zero for $ 4 $ dimensional collapse, but has a nonzero contribution in higher dimensions. We can simplify this equation further to express it in a more familiar form
\begin{equation}\label{isp1}
\frac{A_{0}^{''}}{A_{0}}+(N-3)\frac{B_{0}^{''}}{B_{0}}=\left[\frac{A_{0}^{'}}{A_{0}}+(N-3)\left(\frac{B_{0}^{'}}{B_{0}}\right)\right]\left(\frac{1}{r}+2\frac{B_{0}^{'}}{B_{0}}\right).
\end{equation}
The above relation holds true for $ \Delta=0 $ in a shear-free spherically symmetric spacetime for any dimension $ N\geq4 $.

We need the higher dimensional static solutions of energy density and pressure for our further analysis. They are given by 
\begin{equation}\label{mu0}
\kappa_{N}\mu_{0}=-\dfrac{1}{B_{0}^2}\left[(N-2)\frac{B_{0}^{''}}{B_{0}}+\frac{(N-2)(N-5)}{2}\left(\frac{B_{0}^{'}}{B_{0}}\right)^2+\frac{(N-2)^{2}}{r}\frac{B_{0}^{'}}{B_{0}}\right],
\end{equation}
and
\begin{equation}\label{pres0}
\kappa_{N}p_{0} = \dfrac{1}{B_{0}^2}\left[ \dfrac{(N-2)(N-3)}{2}\Biggl\{\left(\frac{B_{0}^{'}}{B_{0}}\right)^2+ \frac{2}{r}\frac{B_{0}^{'}}{B_{0}}\Biggl\}+(N-2)\Biggl\{\frac{A_{0}^{'}}{A_{0}}\frac{B_{0}^{'}}{B_{0}}+\frac{1}{r}\frac{A_{0}^{'}}{A_{0}} \Biggl\} \right]. 
\end{equation}
Please note that in \eqref{mu0} we get an extra contribution for dimensions $ N>4 $ and the second term inside the parenthesis exactly vanishes for five dimensional collapse.
We can indeed get back these expressions from equations \eqref{pr}, \eqref{mu} by taking $ f $ to be a constant, indicating the initial static state of the collapsing star. 
Also \eqref{qa} shows that for the initial static state the heat flux is zero and with time it increases. (For collapse $ \dot{f} $ is negative, making the heat flux expression positive.)

\subsubsection{Curvature scalars}

We now proceed to derive the curvature scalars in higher dimensions. We start with the conformal Weyl tensor square curvature scalar in $N$ dimensions
\begin{align}\label{NWeylsq}
 \mathcal{W} = \dfrac{4(N-3)}{(N-1)B_{0}^4 f^{4}r^2}\bigg[\frac{A_{0}^{''}}{A_{0}}r-\frac{B_{0}^{''}}{B_{0}}r-2\frac{A_{0}^{'}}{A_{0}}\frac{B_{0}^{'}}{B_{0}}r + 2\bigg(\frac{B_{0}^{'}}{B_{0}}\bigg)^{2}r-\frac{A_{0}^{'}}{A_{0}}+\frac{B_{0}^{'}}{B_{0}}\bigg]^{2} .
\end{align}
Notice the dimensional dependence as $ (N-3)/(N-1)$, which transparently demonstrates that as the dimension of the spacetime increases the magnitude of the conformal Weyl curvature scalar increases. Also the Weyl scalar varies as $ 1/f^{4} $ with time, making it diverge as $ f $ diminishes to zero. Now, using the pressure isotropy condition \eqref{isp1}, we rewrite the Weyl square \eqref{NWeylsq} in the compact form
\begin{align}\label{WeylsqN}
\mathcal{W}=\dfrac{4(N-3)(N-2)^{2}}{(N-1)}(B_{0}f)^{-4}X^2,\,\,X\equiv \left( \frac{B_{0}^{''}}{B_{0}}-2\frac{{B_{0}^{'}}^{2}}{B_{0}^{2}} -\frac{B_{0}^{'}}{rB_{0}}  \right) .
\end{align}
This is the final expression for Weyl curvature scalar in $ N $ dimensions. 
The expression for Weyl curvature scalar in four dimensions matches with results obtained in \cite{Bonnor1}, confirming the validity of the expression for $N$ dimensions.

Now we need to calculate the Ricci scalar $R$. 
After some algebraic simplifications we obtain the Ricci scalar as
\begin{align}\label{Ricci4}
R =&\; \dfrac{1}{f^{2}A_{0}^{2}}\Bigg[-\dfrac{2A_{0}^{''}A_{0}}{B_{0}^{2}}-(2N-4)\dfrac{B_{0}^{''}A_{0}^2}{B_{0}^3}+2(N-1)f\ddot{f} \nonumber \\
& -(N-2)(N-5)\dfrac{A_{0}^{2}{B_{0}^{'}}^{2}}{B_{0}^{4}}-\dfrac{2A_{0}}{rB_{0}^{3}}\left((N-3)rA_{0}^{'}+(N-2)^{2}A_{0}\right)B_{0}^{'} \nonumber \\
& + (N-1)(N-2)\dot{f}^{2}-2(N-2)\dfrac{A_{0}A_{0}^{'}}{rB_{0}^{2}}\Bigg].
\end{align}
In the above equation \eqref{Ricci4} we notice that the fourth term inside the parenthesis vanishes in five dimensional collapse. In this equation we separate the $ \dot{f} $ and $ \ddot{f} $ components from the static components. Using the relations \eqref{fnd}, \eqref{mu0} and\eqref{pres0} we can rewrite the Ricci scalar in a concise manner as 
\begin{align}\label{Riccimup}
R =-\dfrac{2\kappa_{N}}{N-2}T= \dfrac{1}{f^2 A_{0}^2}(N-1)(2a\dot{f}+{\dot{f}}^2)+\dfrac{2\kappa_{N}}{(N-2)f^2}\bigg(\mu_{0}-(N-1)p_{0}\bigg).
\end{align}
In the above expression \eqref{Riccimup}, 
$T$ is the trace of the stress-energy tensor.

Similarly we can calculate the Ricci square scalar $ \mathcal{R} $. We now present a general analytical form in terms of matter variables for any dimension $ N $. Utilizing the trace-reversed field equation we can express the Ricci tensor square in terms of stress-energy tensor and its trace. Consequently, we can express the Ricci tensor square as 
\begin{align}\label{Riccisqgen}
\mathcal{R}= &\;\kappa_{N}^{2}\Bigg[\mu^{2}\left(1-\dfrac{(N-4)}{(N-2)^{2}}\right)+p^{2}\Biggl\{(N-1)-\dfrac{(N-4)(N-1)^{2}}{(N-2)^{2}}\Biggl\}\nonumber\\
&+\dfrac{2(N-1)(N-4)}{(N-2)^{2}}\mu p - 2q^{2}B^{2}\Bigg]=\frac{1}{f^{4}}\left[\sum_{\alpha=0}^{4}\Theta_{\alpha}^{N}(r)\dot{f}^{\alpha}\right].
\end{align}
Here fluid energy density, isotropic pressure and the heat flow are given by \eqref{pr}, \eqref{mu} and \eqref{qa} respectively. The above result \eqref{Riccisqgen} simplifies to the expression obtained in \cite{Bonnor1} in four dimensions. It is visible that the expression is too complicated to be expressed in terms of metric potentials (in a concise manner) for higher dimensions.
The coefficients $ \Theta_{i}^{N} $ in \eqref{Riccisqgen} are functions of $ r $.

Finally we compute the Kretschmann scalar in $ N $ dimensions in the following manner
\begin{equation}\label{krN0}
\mathcal{K}=\frac{4(N-2)}{f^4}\left[\sum_{\alpha=0}^{4}\Omega_{\alpha}^{N}(r)\dot{f}^{\alpha}\right],
\end{equation}
where $\Omega_{\alpha}^{N}$ are functions of $ r $. The expression \eqref{krN0} of the Kretschmann scalar in $N$ dimensions is sufficient for our purpose and the advantage of such a representation will be evident in the next section. The curvature scalars described above are related in $ N $ dimensions by the following
\begin{equation}
\mathcal{K}=\mathcal{W}+\frac{4}{N-2}\mathcal{R}-\frac{2}{(N-1)(N-2)}R^{2}.
\end{equation}

\subsubsection{Epoch functions and arrow of time}

We now compute the gravitational epoch functions and analyse its behaviour. Following the above derived expressions \eqref{WeylsqN} and \eqref{Riccisqgen} for curvature scalars we can deduce the following
\small
\begin{align}\label{BonP}
\mathcal{P}=\dfrac{4(N-3)(N-2)^2}{(N-1)}\dfrac{X^2}{B_{0}^4}\left[\sum_{\alpha=0}^{4}\Theta_{\alpha}^{N}(r)\dot{f}^{\alpha}\right]^{-1}.
\end{align}
\normalsize
It is evident that the ratio of the curvature scalars depend on the dimension. It is interesting to note the behaviour of the epoch function with time during gravitational collapse. For this we need \eqref{dotfN} to understand the overall behaviour of $\mathcal{P}$. In the beginning the integration constants can be chosen in such a manner that the temporal function $ f $ becomes a constant. Without any loss of generality, this constant can be assumed to be of unit value. Consequently, in this limit, the epoch function $ \mathcal{P} $ is a positive quantity. However, as time passes the value of $ f $ tends to zero with $ t\rightarrow t_{s} $, making the time derivative of $ f $ to diverge. As a result in this limit, as the collapse proceeds, the epoch function vanishes. This means that the gravitational entropy due to $ \mathcal{P} $ decreases with time during collapse even in higher dimensions.

Now we investigate the epoch function $ \mathcal{P}_{1} $ for higher dimensions. Taking the ratio of the Weyl square \eqref{WeylsqN} and Kretschmann scalar \eqref{krN0} we get $ \mathcal{P}_{1} $ as
\small
\begin{align}\label{RudP}
\mathcal{P}_{1}= \dfrac{(N-3)(N-2)}{(N-1)}\dfrac{X^2}{B_{0}^4}\left[\sum_{\alpha=0}^{4}\Omega_{\alpha}^{N}(r)\dot{f}^{\alpha}\right]^{-1}.
\end{align}
\normalsize
In this case too when we substitute $ \dot{f} $ using the first integral \eqref{dotfN} into appropriate limits we find that initially the epoch function $ \mathcal{P}_{1} $ starts from a positive value and as time progresses it goes to zero. This means that even for this epoch function, the gravitational entropy decreases with time in collapse. Consequently, the arrow of time associated with both of these gravitational epoch functions (gravitational arrow of time) are in the opposite direction of the the arrow of time associated with the outgoing radiation (thermodynamic arrow of time).

\subsection{The second spacetime: Time-modulated lapse function}
In this section we consider another interesting case for the interior spacetime metric when the lapse function $ A(r,t)=A_{0}(r)f(t) $ is time-dependent and separable. As a result the interior metric can be written in the following manner
\begin{equation}
ds_{-}^2=f(t)^{2}\left\{-A_{0}(r)^2 dt^2+B_{0}(r)^2 (dr^2+r^{2}d\Omega_{N-2}^{2})\right\}.
\end{equation}
The resulting metric is conformal to the shear-free static spacetime. By solving the Einstein's field equations in $N$ dimensions we obtain the matter variables
\begin{equation}\label{prt}
p=\frac{p_{0}}{f^2}-\frac{1}{\kappa_{N}A_{0}^2}\bigg\lbrace(N-2)\frac{\ddot{f}}{f}+\dfrac{(N-2)(N-5)}{2}\left(\frac{\dot{f}}{f}\right)^2\bigg\rbrace \dfrac{1}{f^2} ,
\end{equation}
\begin{equation}\label{mut}
\mu=\frac{\mu_{0}}{f^2}+\frac{(N-1)(N-2)}{2\kappa_{N} f^{2} A_{0}^2}\left(\frac{\dot{f}}{f}\right)^2 ,
\end{equation}
\begin{equation}\label{qat}
q^{\alpha}=q\delta_{1}^{\alpha}=-\dfrac{(N-2)A_{0}^{'}}{\kappa_{N} A_{0}^{2}B_{0}^{2}}\left(\frac{\dot{f}}{f^{4}}\right)\delta_{1}^{\alpha} .
\end{equation}
It is important to notice the difference from our previous case, especially in the pressure expression the second term inside the parenthesis vanishes in five dimensions. Moreover, in each of the above expressions we have extra contributions of $ 1/f $.
Therefore, the matching condition at the boundary hypersurface $ \Sigma $ can be reduced to the following form
\begin{equation}\label{modODE}
2f\ddot{f}+(N-5){\dot{f}}^{2}-2af\dot{f}=0.
\end{equation}
Subsequently, the first integral of the above differential equation can be obtained as
\begin{equation}\label{moddotf}
\dfrac{\dot{f}}{f}=\dfrac{2a}{(N-3)}+Cf^{-(N-3)/2}.
\end{equation}
By integrating the above equation we get the explicit form of $ f $ as
\begin{equation}\label{lapsef}
f(t)=\left[\dfrac{1}{2a}{(N-3)(e^{a(t+t_{0})}-C)}\right]^{2/(N-3)},
\end{equation}
where $ C $ and $ t_{0} $ are the constants of integration.

From the field equations it follows that the condition of pressure isotropy remains unchanged. Subsequently, a similar analysis can be followed in this case and the curvature scalars obtained. We will not mention the scalars explicitly for brevity of this paper; however without going into the detailed higher dimensional functional forms we can observe some temporal patterns. After a long algebraic simplification we can reduce the epoch function $ \mathcal{P} $ to
\begin{equation}
\mathcal{P}\sim \left[\sum_{\alpha=0}^{4}\Xi_{\alpha}^{N}(r) {\left({\dot{f}}/{f}\right)}^{\alpha}\right]^{-1},
\end{equation}
where the denominator is a fourth order polynomial of the first integral $ \dot{f}/f $ with $ \Xi_{i}^{N}(r) $s as coefficients. In a similar manner the epoch function $ \mathcal{P}_{1} $ can also be deduced and expressed in the following manner
\begin{equation}
\mathcal{P}_{1}\sim \left[\sum_{\alpha=0}^{4}\Upsilon_{\alpha}^{N}(r) {\left({\dot{f}}/{f}\right)}^{\alpha}\right]^{-1}.
\end{equation}
Here also the denominator has the same pattern with coefficients $ \Upsilon_{i}^{N}(r) $. In these coefficients the superscript $ N $ denotes the dimensionality, and the subscript $ i $ denotes the order of the coefficient in the polynomial. Now we are in a position to analyse the behaviour of the epoch functions with time evolution. From the first integral \eqref{moddotf} it is clear that $ \dot{f}/f $ diverges to infinity as $ f $ decreases to zero with time. As a result the denominator in the epoch functions diverges as time progresses starting from a finite value. This conclusively demonstrates that the gravitational entropy decreases with collapse and the arrow of time due to gravitational entropy is opposite to the thermodynamic arrow of time associated with radiation.
\section{Arrow of time problem in presence of charge}
We now introduce charge in the interior matter distribution and see how it affects our previous results. The idea here is to check whether the directionality of gravitational arrow of time changes due to the effect of charge in the system. Several works are available in literature on the charged matter distributions in \cite{[14],[15],[16],olivera2}. For this study we will again consider the shear-free interior spacetime \eqref{int_soln}, and the energy momentum tensor for the interior matter distribution gets modified by the electromagnetic energy tensor $ E_{ab} $ to give
\begin{equation}
T_{ab}=(\mu + p)v_{a}v_{b}+p \textsl{g}_{ab}+q_{a}v_{b}+q_{b}v_{a}+\pi_{ab}+E_{ab}.
\end{equation}
The electromagnetic energy tensor $ E_{ab} $ is defined in terms of Faraday tensor $F_{ab}$ as 
\begin{equation}
E_{ab}=\frac{1}{\mathcal{A}_{N-2}}\left(F_{a}^{\,\,c}F_{bc}-\frac{1}{4}F^{cd}F_{cd} \textsl{g}_{ab}\right),\,\,\,\,\mathcal{A}_{N-2}=\frac{2\pi^{\frac{N-1}{2}}}{\Gamma\left(\frac{N-1}{2}\right)},
\end{equation}
where $\mathcal{A}_{N-2}$ is the surface area of the $ (N-2)$-sphere and $ \Gamma(...)$ is the gamma function. The Maxwell bivector or Faraday tensor is defined as 
\begin{equation}
F_{ab}=-F_{ba}=\Phi_{b;a}-\Phi_{a;b},
\end{equation}
with $ \Phi_{a} $ as the electromagnetic potential, which can be denoted by $ \Phi_{a}=(\phi(r,t),0,0,.\,.\,.\,,0) $. In the presence of charge we consider the Einstein-Maxwell field equations as the following
\begin{eqnarray}
G_{ab}=\kappa_{N}T_{ab},\,\, F_{[ab;c]}=0,\,\,\,F_{\,\,\,\,\,\,;b}^{ab}=\mathcal{A}_{N-2}J^{a}.
\end{eqnarray}
Here $J^{a}$ is the current defined as $ J^{a}=\zeta u^{a} $ with $ \zeta $ as the proper charge density. From these sets of equations we derive the field equations for the electromagnetic field.
\begin{align}
\phi^{''}-\left[\frac{A^{'}}{A}+\frac{B^{'}}{B}-(N-2)\frac{Y^{'}}{Y}\right]\phi^{'}=\mathcal{A}_{N-2}\zeta A B^{2}, \label{phi1}\\
{\dot{\phi}}^{'}-\left(\frac{\dot{A}}{A}+\frac{\dot{B}}{B}-(N-2)\frac{\dot{Y}}{Y}\right)\phi^{'}=0. \label{phi2}
\end{align}
From equations \eqref{phi1} and \eqref{phi2} we get the total conserved charge of the star as a function of $ r $ as
\begin{equation}
Q(r)=\mathcal{A}_{N-2}\int^{r}\zeta B Y^{N-2}d\tilde{r}.
\end{equation}

We will now only consider the first spacetime \eqref{int_soln} from the previous section and introduce charge.
Following the previous treatment but now including charge we arrive at the following differential equation for $ f $ as
\begin{equation}\label{ffQ}
2f\ddot{f}+(N-3)\dot{f}^2-2\left(\frac{A_{0}^{'}}{B_{0}}\right)_{\Sigma} \dot{f}+\frac{\kappa_{N}}{(N-2)\mathcal{A}_{N-2}}\left(\frac{A_{0}^{2}Q^2}{(rB_{0})^{2N-4}}\right)_{\Sigma}=0.
\end{equation}
Let us denote the last term in the LHS as 
\begin{equation}
\varsigma(r_{\Sigma})=\frac{\kappa_{N}}{(N-2)\mathcal{A}_{N-2}}\left(\frac{A_{0}^{2}Q^2}{(rB_{0})^{2N-4}}\right)_{\Sigma}.
\end{equation}
{\renewcommand{\arraystretch}{3.5}
\begin{table}[hbt!]
   \centering
   \rotatebox{0}{
   \begin{minipage}{\textwidth}
   \centering
   \caption{Nature of the differential equation for $ f $ in higher dimensions with charge $Q$} 
    \label{tabf2}
    \resizebox{\textwidth}{!}{
   \begin{tabular}{lcr}
   \toprule\toprule
   {\Large\textbf{$\Delta$}}  & {\Large\textbf{First integral}} & {\Large\textbf{Second integral}} \\ 
   \midrule
   $>0$ & $\!\begin{aligned}[t]&\vert\dot{f}-v_{1}\vert^{A}\vert\dot{f}-v_{2}\vert^{B}=e^{C_{1}}f;\\
  &A=-\frac{2v_{1}}{(N-3)(v_{1}-v_{2})},\\
  &B= -\frac{2v_{2}}{(N-3)(v_{2}-v_{1})}, \\
  &v_{1,2}=\frac{a\pm\sqrt{a^{2}-(N-3)\varsigma(r_{\Sigma})}}{N-3}.
   \end{aligned}$ & $\!\begin{aligned}[t] & t=\int \frac{df}{v}+C_{2}; \\   
   & \text{where v is implicitly defined by}\\
   & \int \frac{2vdv}{-(N-3)v^{2}+2av-\varsigma(r_{\Sigma})}=\ln f+C_{1}
   \end{aligned}
   $  \\
   $=0$ & $\!\begin{aligned}[t]&-\frac{2}{(N-3)}\ln\left\vert\dot{f}-\frac{a}{N-3}\right\vert+\frac{2a}{(N-3)^{2}\left(\dot{f}-\frac{a}{N-3}\right)}\\
   &=\ln f+C_{1} \end{aligned}$  & $ \!\begin{aligned}[t]
   &t+C_{2}=\\
   & \int \frac{df}{\frac{a}{N-3}\left(1+\frac{1}{\ln \left\vert \dot{f}-\frac{a}{N-3}\right\vert+\frac{(N-3)}{2}\ln f+\frac{(N-3)}{2}C_{1}}\right)}; \end{aligned}$ \\
   $<0$ & $\!\begin{aligned}[t]&\ln f+C_{1}=-\frac{1}{N-3}\ln\left[\left(\dot{f}-\frac{a}{N-3}\right)^{2}+Q\right]\\
   & -\frac{a}{(N-3)\sqrt{(N-3)\varsigma(r_{\Sigma})-a^{2}}}\\
   &\arctan\left[\frac{(N-3)\dot{f}-a}{\sqrt{(N-3)\varsigma(r_{\Sigma})-a^2}}\right];\\
   & Q=  \frac{(N-3)\varsigma(r_{\Sigma})-a^2}{(N-3)^2},Q>0     
   \end{aligned}$ & $\!\begin{aligned}[t] &  t=-\int\frac{N-3}{2}\frac{u^2+Q}{u+\frac{a}{N-3}}d(\ln f+C_{1})+C_{2}; \\
   & u=\dot{f}-\frac{a}{N-3}
   \end{aligned}$ \\
   \bottomrule
   \end{tabular}
   }
   \end{minipage}}
\end{table}
}
To solve equation \eqref{ffQ}, we make the substitution $ v=\dot{f} $, and simplify to get the following
\begin{equation}\label{intQ}
\int \frac{2v dv}{-(N-3)v^2 +2av-\varsigma(r_{\Sigma})}=\ln f +C_{1},
\end{equation}
where $ C_{1} $ is the constant of integration and $ a=(A_{0}^{'}/B_{0})_{\Sigma} $.
It is evident that the nature of the solution of equation  \eqref{intQ} depends on the quadratic term $ -(N-3)v^2 +2av-\varsigma(r_{\Sigma}) $. Therefore we compute the discriminant of the quadratic expression, which is given below 
\begin{equation}
\Delta=4[a^{2}-(N-3)\varsigma(r_{\Sigma})].
\end{equation}
In Table \ref{tabf2} we present the first and second integrals of \eqref{ffQ} which depends on the sign of discriminant $ \Delta $. 

\subsection{Analysis of $ f $ and consequences on arrow of time}

A careful analysis of each case is necessary to understand how these solutions behave and whether they are physically viable. For $ \Delta >0 $ the first integral contains the different parameters $ A, B $ and $C_{1}$. From the expressions of $ v_{1} $ and $v_{2}$ it is evident that both of them are positive as $ \varsigma(r_{\Sigma})>0 $. The integration constant $ C_{1}$ can be fixed by imposing the initial conditions $ f=1 $ and $ \dot{f}\approx 0 $. Moreover, it is evident that $ v_{1}>v_{2} $, which makes $ A<0 $ and $ B>0 $. Now let us consider the case where $ f\rightarrow 0 $. In this limit one might think that it is only possible for $ \dot{f} \rightarrow v_{1},v_{2} $, but in a collapsing situation $ \dot{f} $ is negative and therefore these cannot be the desired solutions. The only viable solution is $ \dot{f}\rightarrow -\infty  $ when $ f\rightarrow 0 $, because in this limit the LHS of the first integral is dominated by the first term, making the entire expression diverge to negative infinity. Therefore, when $ \Delta >0 $, the only possible and physically viable solution is $ f=1 $ with vanishing $ \dot{f} $ in the initial phase and in the end state $ f\rightarrow 0 $ with $ \dot{f}\rightarrow -\infty $. Now let us consider the case for $ \Delta=0 $. We start by taking $ f=1 $ and $ \dot{f}\approx 0 $ to fix the integration constant $ C_{1} $. Then we consider the limit $ f\rightarrow 0 $ and try to find a viable solution for $ \dot{f} $. As $ \dot{f}<0 $, it cannot take the value $ a/(N-3) $ and also it makes the LHS diverge to positive infinity unlike the RHS. Therefore, the only possible solution for $ \dot{f} $ is when it diverges to negative infinity. So in this case also the viable solution of $ f $ remains the same. Finally, we consider the case for $ \Delta<0 $. We can fix the integration constant $ C_{1} $ by taking $ f=1 $ and $ \dot{f}\approx 0 $. Considering the end state of the collapse we take $ f\rightarrow 0 $. This choice leaves us with the only possible solution for RHS as $ \dot{f}\rightarrow -\infty $, making the physically viable option as the only solution.

Consequently, in presence of charge, shear-free collapse with a generalised Vaidya exterior has viable solutions as identified in Table \ref{tabf2}. 

We will now proceed to investigate the nature of the gravitational epoch functions in such a system. Here the Einstein tensor is of the form \eqref{shfrein}. Therefore, we can use the epoch functions in \eqref{BonP} and \eqref{RudP}. It has already been discussed that the nature of these epoch functions depend on how the function  $\dot{f}$ behaves. From the above analysis it is evident that the only possible physically viable solution is when $ \dot{f} $ starts from zero and decreases rapidly towards negative infinity as the collapse progresses. This means that both the epoch functions start with a positive definite value and with collapse their value diminishes to zero, resulting in a decreasing gravitational entropy. As a result, the arrow of time associated with it is in the opposite direction of the thermodynamic arrow of time.

\subsection{Charged Vaidya exterior}

We can also consider a special case of the generalized Vaidya metric with the following mass function
\begin{equation}
m(u,R)=M(u)-\dfrac{\kappa_{N}Q^{2}}{2(N-2)\mathcal{A}_{N-2}R^{N-3}},
\end{equation}
with 
\begin{equation}
M(u)=\left[\left(\dfrac{N-3}{2}\right)Y^{N-3}\left(1+\dfrac{{\dot{Y}}^2}{A^2}-\dfrac{{Y^{'}}^2}{B^2}\right)+\dfrac{\kappa_{N}Q^{2}}{2(N-2)\mathcal{A}_{N-2}R^{N-3}}\right]_{\Sigma}.
\end{equation}
This mass function simplifies the generalised Vaidya metric into the following form
\begin{align}\label{extQ}
ds_{+}^2= & -\left(1-\frac{2M(u)}{(N-3)R^{N-3}}+\dfrac{\kappa_{N}Q^{2}}{(N-2)(N-3)\mathcal{A}_{N-2}R^{2(N-3)}}\right)du^2 \nonumber \\
 & -2dudR+ R^2d\Omega_{N-2}^{2}.
\end{align}
This is the metric for the higher dimensional charged Vaidya spacetime. In this case the matching condition at the boundary hypersurface $ \Sigma $ remains the same as \eqref{ffQ} and therefore, the corresponding dynamics of $ f(t) $ are the same as in Table \ref{tabf2}. Interestingly in the presence of null strings the matching condition simplifies further as the energy density of the null string exactly cancels the contribution of the charge present \cite{byron} and the matching condition follows the differential equation \eqref{fnd} and the interior dynamics follows Table \ref{tabf}. As a result in this case also when a null string is present the gravitational arrow of time is opposite to that of the radiation. In fact for a general composite interior matter distribution the behaviour of the gravitational arrow of time remains same. 

\section{Discussion and conclusions}

In this paper $ N $ dimensional shear-free dissipative spherical collapse was discussed in detail with both unmodulated and modulated lapse function. The charged collapse was also taken into account in the latter portion of the work. To keep the calculations simple we neglected pressure anisotropy and derived the pressure isotropy relation for $ N $ dimensions and showed how it deviates in the presence of charge. For both (uncharged and charged) systems we derived the resulting differential equation of $ f $ on the $N-1$ dimensional boundary hypersurface $ \Sigma $. In both the cases a class of solutions of $ f $ were found and presented in tabular form. The higher dimensional curvature scalars were computed and their functional behaviour in terms of matter variables were determined wherever possible. We deduced the gravitational epoch functions for both the cases and found that they decrease with collapse. Consequently, the gravitational arrow of time is opposite to the thermodynamic arrow of time for every dimension $ N\geq4 $.

It is important to note the following:

\begin{itemize}
\item  The alignment of the gravitational arrow of time with respect to the thermodynamic time arrow doesn't change with spatial dimensions $ N $.

\item It is important to note that we have only considered the shear-free class of solutions in our study. The effect of shear in gravitational collapse was studied before in \cite{naked}, where it was shown that a sufficiently strong shearing effect can delay the formation of horizon, leading to a naked singularity. Also in \cite{shear} it was concluded that in a shear-free collapse the end state always leads to a black hole. In absence of shear (respecting the usual energy conditions), no naked singularity can form as an end state of collapse. This implies, in a shear-free situation the Weyl curvature can never diverge faster than the Ricci curvature. As a result,
the gravitational epoch functions $\mathcal{P}$ and $\mathcal{P}_{1}$ will always decrease. Consequently, in such a system the arrow of time will always be in the opposite direction. Hence our current study has validated, and is in conformity with the previously obtained results \cite{ndimcollap}, and indicates the important role of shear in the arrow of time problem.
\item We know that (in a more general setting) the evolution of shear is related to the anisotropic stress. Therefore, it follows from the previous argument that the collapse end state is related to the complexity of the system. Also, the behaviour of the curvature scalars are important because it not only helps to estimate the gravitational arrow of time but also related to the complexity. It was shown by Herrera et al. \cite{split} that the orthogonal splitting of the Riemann tensor and its double dual with respect to the fluid velocity gives the structure scalars of the system
\begin{equation}
Y_{ab}=R_{acbd}v^{c}v^{d} \,\,\,\,\,\, X_{ab}={}^{*}R_{acbd}^{*}v^{c}v^{d}.
\end{equation}
As the system is spherically symmetric, the magnetic part of the Weyl tensor $ H_{ab}=0 $. Therefore, the electric part of the Weyl tensor $(E_{ab})$ is sufficient to express the Weyl tensor as
\begin{equation}
E_{ab}=C_{acbd}v^{c}v^{d}=\mathscr{E} \left(\mathcal{N}_{a}\mathcal{N}_{b}-\dfrac{1}{N-1}h_{ab}\right).
\end{equation}
In four dimensions the complexity factor $ Y_{TF} $, the trace-free part of the tensor $ Y_{ab} $, can be obtained as
\begin{equation}
Y_{TF}=\mathscr{E}-4\pi\Delta.
\end{equation}
Now we know that $ C_{abcd}C^{abcd} $ is basically $ \mathscr{E}^2 $, which means that the structure scalar $ Y_{TF} $ (complexity factor) is directly related to the gravitational epoch function and the pressure anisotropy $ (\Delta) $. For our case where we have assumed the pressure anisotropy to be zero, the gravitational epoch function is directly proportional to the complexity of the system. It is important to note that gravitational entropy is proportional to the gravitational epoch function. Therefore the conclusion remains same when we measure the gravitational entropy in the CET proposal \cite{CET}, where the gravitational energy density is proportional to the Weyl scalar (in Newman-Penrose formalism) $ \vert\Psi_{2}\vert $. Therefore, we can conclude the following 

\begin{itemize}
\item In spherical symmetry (especially in LRS II spacetimes) the gravitational entropy is proportional to the complexity factor in a system with pressure isotropy.
\item In spherical symmetry (especially in LRS II spacetimes) the gravitational entropy is proportional to the complexity factor and the pressure anisotropy where $ \Delta\neq0 $.
\end{itemize}

As the arrow of time and gravitational entropy are linked intimately, we can argue that there is a deep connection between the complexity and the arrow of time. 

\item Essentially the entire study is a local application of WCH in higher dimensions. We have indeed found that the Weyl curvature scalar derived functions like $ \mathcal{P} $ and $ \mathcal{P}_{1} $, give a proper measure of gravitational epoch function, however, they fail to give a properly oriented local gravitational arrow of time. This means that it is not sufficient to have a well behaved gravitational epoch function for WCH to hold locally, the proper orientation of the arrow of time is necessary. 
\item 
Determination of the temporal function $f(t)$ is very important as it not only controls the gravitational arrow of time but also governs many physical features of a radiating star apart from the matching at $ \Sigma $.
The horizon function $H$ was introduced by Ivanov and others in \cite{[52],[53]}. For shear-free collapse the function $ H $ is of the following form
\begin{equation}
H=1+\dfrac{rB_{0}^{'}}{B_{0}}+\dfrac{r B_{0}\dot{f}}{A}.
\end{equation}
It is evident that by determining the functional forms of $ f $ we can fix the temporal part of $ H $. The horizon function is useful in the determination of various stellar features like the mass, energy density, the redshift $ z_{\Sigma} $, surface luminosity $ \Lambda_{\Sigma} $ and the luminosity at infinity $ \Lambda_{\infty} $. It is evident that the orientation of the arrow of time is related to the function $ H $, via $ \dot{f} $, hence it is deeply connected to various physical processes.

\item
Moreover, for the determination of temperature profiles in causal thermodynamics $ f $ is very important. The Maxwell-Cattaneo equation in a shear-free spherically symmetric spacetime becomes
\begin{equation}
(f^{2}q)^{\dot{•}}T^{-\sigma}f+\left(\dfrac{Aq}{\beta}\right)f^{2}=-\dfrac{\alpha T^{3-\sigma}}{\beta B_{0}^{3}}(AT)^{'},
\end{equation}
where all the symbols follow their usual conventional meanings. For more details please see \cite{[54],[55],[56],[57],[58]}. For the functional forms found for $ f $ we can solve the above equation to get the temperature profiles of the radiating star. A detailed study of this equation can also be found in \cite{[56]}.
This relation establishes the connection between the arrow of time and causal thermodynamics via $ f $ and $ \dot{f} $.
\end{itemize}
 This study indicates that we need more detailed studies on the local application of Weyl curvature hypothesis. 

\section*{Acknowledgements}
SC is thankful to the University of KwaZulu-Natal (UKZN) for post doctoral funding. RG thanks UKZN and NRF, South Africa, for support. 
SG thanks UKZN for the visiting professorship and IUCAA, India, for an associateship.

\section*{Data availability statement}
All data that support the findings of this study are included within the article (and any supplementary files). 

\section*{Code availability statement}
There is no code associated with this article.
\section*{Conflict of interest}
We confirm no conflicts of interest.


\begin{thebibliography}{999}

\bibitem{Penrose1} 
Penrose, R. Spacetime singularities. In \textit{Proceedings of the First Marcel Grossmann Meeting on General Relativity}; Ruffini, R., Ed.; North-Holland: Amsterdam, The Netherlands, 1977; pp. 173-181.

\bibitem{Penrose2} 
Penrose, R. Difficulties with inflationary cosmology. \textit{Ann. N. Y. Acad. Sci.} \textbf{1989}, \textit{571}, 249-264.

\bibitem{Bolejko} 
Bolejko, K. Gravitational entropy and the cosmological no-hair conjecture. \textit{Phys. Rev. D} \textbf{2018}, \textit{97}, 083515.

\bibitem{Bonnor1} 
Bonnor, W.B. The gravitational arrow of time. \textit{Phys. Lett. A} \textbf{1985}, \textit{112}, 26-28. 

\bibitem{Bonnor2} 
Bonnor, W.B. The gravitational arrow of time and the Szekeres cosmological models. \textit{Class. Quantum Grav.} \textbf{1986}, \textit{3}, 495-501. 

\bibitem{Bonnor3} 
Bonnor, W.B. Arrow of time for a collapsing, radiating sphere. \textit{Phys. Lett. A} \textbf{1987}, \textit{122}, 305-308. 

\bibitem{Arrow1} 
Chakraborty, S.; Maharaj, S.D.; Guha, S.; Goswami, R. Arrow of time and gravitational entropy in collapse. \textit{Class. Quantum Grav.} \textbf{2024}, \textit{41}, 127003. 

\bibitem{entropy1} 
Rudjord, Ø.; Grøn, Ø.; Hervik, S. The Weyl curvature conjecture and black hole entropy. \textit{Phys. Scr.} \textbf{2008}, \textit{77}, 055901. 

\bibitem{entropy2} 
Romero, G.E.; Thomas, R.; Pérez, D. Gravitational entropy of black holes and wormholes. \textit{Int. J. Theor. Phys.} \textbf{2012}, \textit{51}, 925-942. 

\bibitem{Bekenstein} 
Bekenstein, J.D. Black holes and entropy. \textit{Phys. Rev. D} \textbf{1973}, \textit{7}, 2333-2346. 

\bibitem{SWH1} 
Hawking, S.W. Particle creation by black holes. \textit{Commun. Math. Phys.} \textbf{1975}, \textit{43}, 199-220. 

\bibitem{Chandra} 
Chandrasekhar, S. \textit{The Mathematical Theory of Black Holes}; Oxford University Press: Oxford, UK, 1983.

\bibitem{gc} 
Guha, S.; Chakraborty, S. On the gravitational entropy of accelerating black holes. \textit{Int. J. Mod. Phys. D} \textbf{2020}, \textit{29}, 2050034. 

\bibitem{cgg} 
Chakraborty, S.; Guha, S.; Goswami, R. An investigation on gravitational entropy of cosmological models. \textit{Int. J. Mod. Phys. D} \textbf{2021}, \textit{30}, 2150051. 

\bibitem{cgg1} 
Chakraborty, S.; Guha, S.; Goswami, R. How appropriate are the gravitational entropy proposals for traversable wormholes? \textit{Gen. Relativ. Gravit.} \textbf{2022}, \textit{54}, 47. 

\bibitem{CET} 
Clifton, T.; Ellis, G.F.R.; Tavakol, R. A gravitational entropy proposal. \textit{Class. Quantum Grav.} \textbf{2013}, \textit{30}, 125009. 

\bibitem{misha} 
Gromov, M. Four lectures on scalar curvature. \textit{arXiv} \textbf{2019}, arXiv:1908.10612 [math.DG].

\bibitem{[1]} 
Emel'yanov, V.M.; Nikitin, Y.P.; Rozental, I.L.; Berkov, A.V. Physics in multidimensional spaces and the beginning of metagalaxy. \textit{Phys. Rep.} \textbf{1986}, \textit{143}, 1-68. 

\bibitem{[2]} 
Tangherlini, F.R. Schwarzschild field in n dimensions and the dimensionality of space problem. \textit{Nuovo Cim.} \textbf{1963}, \textit{27}, 636-651. 

\bibitem{[3]} 
Myers, R.C.; Perry, M.J. Black holes in higher dimensional spacetimes. \textit{Ann. Phys.} \textbf{1986}, \textit{172}, 304-347. 

\bibitem{[4]} 
Dianyan, X. Exact solutions of Einstein and Einstein-Maxwell equations in higher-dimensional spacetime. \textit{Class. Quantum Grav.} \textbf{1988}, \textit{5}, 871. 

\bibitem{[5]} 
Paul, B.C. Relativistic star solutions in higher dimensions. \textit{Int. J. Mod. Phys. D} \textbf{2004}, \textit{13}, 229-238. 

\bibitem{[6]} 
Andrade, T.; Emparan, R.; Licht, D.; Luna, R. Black hole collisions, instabilities, and cosmic censorship violation at large D. \textit{J. High Energy Phys.} \textbf{2019}, \textit{09}, 099. 

\bibitem{[7]} 
Ponce de Leon, J.; Cruz, N. Hydrostatic equilibrium of a perfect fluid sphere with exterior higher-dimensional Schwarzschild spacetime. \textit{Gen. Relativ. Gravit.} \textbf{2000}, \textit{32}, 1207-1216. 

\bibitem{[8]} 
Wright, M. Buchdahl type inequalities in d dimensions. \textit{Class. Quantum Grav.} \textbf{2015}, \textit{32}, 215005. 

\bibitem{[9]} 
Harko, T.; Mak, M.K. Anisotropic charged fluid spheres in D
 space–time dimensions. \textit{J. Math. Phys.} \textbf{2000}, \textit{41}, 4752-4764. 
 
\bibitem{[10]} 
Wang, C.; Xu, Z.-M.; Wu, B. Mass-radius ratio bound for horizonless charged compact object in higher dimensions. \textit{Phys. Lett. B} \textbf{2020}, \textit{802}, 135234. 

\bibitem{[11]} 
Santos, N.O. Non-adiabatic radiating collapse. \textit{Mon. Not. R. Astron. Soc.} \textbf{1985}, \textit{216}, 403-410. 

\bibitem{[12]} 
Maharaj, S.D.; Govender, G.; Govender, M. Radiating stars with generalised Vaidya atmospheres. \textit{Gen. Relativ. Gravit.} \textbf{2012}, \textit{44}, 1089-1099. 

\bibitem{[13]} 
Maharaj, S.D.; Brassel, B.P. Radiating stars with composite matter distributions. \textit{Eur. Phys. J. C} \textbf{2021}, \textit{81}, 366. 

\bibitem{[14]} 
Maharaj, S.D.; Brassel, B.P. Radiating composite stars with electromagnetic fields. \textit{Eur. Phys. J. C} \textbf{2021}, \textit{81}, 783. 

\bibitem{[15]} 
Tikekar, R.; Patel, L.K. Non-adiabatic gravitational collapse of charged radiating fluid spheres. \textit{Pramana} \textbf{1992}, \textit{39}, 17-25. 

\bibitem{[16]} 
Banerjee, A.; Dutta Choudhury, S.B. Junction conditions at the boundary of a charged viscous-fluid sphere. \textit{Gen. Relativ. Gravit.} \textbf{1989}, \textit{21}, 785-795. 

\bibitem{[17]} 
Maharaj, S.D.; Govender, M. Collapse of a charged radiating star with shear. \textit{Pramana} \textbf{2000}, \textit{54}, 715-727. 

\bibitem{[18]} 
Bhui, B.; Chatterjee, S.; Banerjee, A. Non-adiabatic gravitational collapse in higher dimensional space-time and its junctions conditions. \textit{Astrophys. Space Sci.} \textbf{1995}, \textit{226}, 7-18. 

\bibitem{[19]} 
Shah, H.; Ahmed, Z.; Khan, S. Higher dimensional shear-free radiating collapse. \textit{Can. J. Phys.} \textbf{2018}, \textit{96}, 1201-1204. 

\bibitem{[20]} 
Banerjee, A.; Chatterjee, S. Spherical collapse of a heat conducting fluid in higher dimensions without horizon. \textit{Astrophys. Space Sci.} \textbf{2005}, \textit{299}, 219-225. 

\bibitem{[21]} 
Iyer, B.R.; Vishveshwara, C.V. The Vaidya solution in higher dimensions. \textit{Pramana} \textbf{1989}, \textit{32}, 749-752. 

\bibitem{[22]} 
Ghosh, S.G.; Dawood, A.K. Radiating black hole solutions in arbitrary dimensions. \textit{Gen. Relativ. Gravit.} \textbf{2008}, \textit{40}, 9-21. 

\bibitem{[23]} 
Chatterjee, S.; Bhui, B.; Banerjee, A. Higher-dimensional Vaidya metric with an electromagnetic field. \textit{J. Math. Phys.} \textbf{1990}, \textit{31}, 2208-2210. 

\bibitem{[24]} 
Ghosh, S.G.; Dadhich, N.K. Naked singularities in higher dimensional Vaidya spacetimes. \textit{Phys. Rev. D} \textbf{2001}, \textit{64}, 047501. 

\bibitem{[25]} 
Ghosh, S.G.; Dadhich, N. Gravitational collapse of type II fluid in higher dimensional spacetimes. \textit{Phys. Rev. D} \textbf{2002}, \textit{65}, 127502. 

\bibitem{[26]} 
Patil, K.D. Gravitational collapse in higher-dimensional charged-Vaidya spacetime. \textit{Pramana} \textbf{2003}, \textit{60}, 423-431. 

\bibitem{[27]} 
Harko, T. Gravitational collapse of a Hagedorn fluid in Vaidya geometry. \textit{Phys. Rev. D} \textbf{2003}, \textit{68}, 064005. 

\bibitem{[28]} 
Saa, A. $N$-dimensional Vaidya metric with a cosmological constant in double-null coordinates. \textit{Phys. Rev. D} \textbf{2007}, \textit{75}, 124019. 

\bibitem{[29]} 
Debnath, U.; Chakraborty, N.C.; Chakraborty, S. Gravitational collapse in higher dimensional Husain space–time. \textit{Gen. Relativ. Gravit.} \textbf{2008}, \textit{40}, 749-763. 

\bibitem{[30]} 
Goswami, R.; Joshi, P.S. Spherical gravitational collapse in $N$ dimensions. \textit{Phys. Rev. D} \textbf{2007}, \textit{76}, 084026. 

\bibitem{[31]} 
Mkenyeleye, M.D.; Goswami, R.; Maharaj, S.D. Is cosmic censorship restored in higher dimensions?. \textit{Phys. Rev. D} \textbf{2015}, \textit{92}, 024041. 

\bibitem{[32]} 
Chatterjee, S.; Ganguli, S.; Virmani, A. Charged Vaidya solution satisfies weak energy condition. \textit{Gen. Relativ. Gravit.} \textbf{2016}, \textit{48}, 91. 

\bibitem{[48]} 
Ghosh, S.G.; Jhingan, S.; Deshkar, D.W. Spherical gravitational collapse in 5D Einstein-Gauss-Bonnet gravity. \textit{J. Phys. Conf. Ser.} \textbf{2014}, \textit{484}, 012013. 

\bibitem{[33]} 
Debnath, U. Thermodynamics in higher dimensional Vaidya Spacetime. \textit{Int. J. Theor. Phys.} \textbf{2014}, \textit{53}, 2108-2117. 

\bibitem{[34]} 
Kobayashi, T. A Vaidya-type radiating solution in Einstein-Gauss-Bonnet gravity and its application to braneworld. \textit{Gen. Relativ. Gravit.} \textbf{2005}, \textit{37}, 1869-1876. 

\bibitem{[35]} 
Dominguez, A.E.; Gallo, E. Radiating black hole solutions in Einstein-Gauss-Bonnet gravity. \textit{Phys. Rev. D} \textbf{2006}, \textit{73}, 064018. 

\bibitem{[36]} 
Brassel, B.P.; Maharaj, S.D.; Goswami, R. Extended naked conical singularity in radiation collapse in Einstein-Gauss-Bonnet gravity. \textit{Phys. Rev. D} \textbf{2018}, \textit{98}, 064013. 

\bibitem{[37]} 
Brassel, B.P.; Maharaj, S.D.; Goswami, R. Higher-dimensional radiating black holes in Einstein-Gauss-Bonnet gravity. \textit{Phys. Rev. D} \textbf{2019}, \textit{100}, 024001. 

\bibitem{[38]} 
Cai, R.-G.; Cao, L.-M.; Hu, Y.-P.; Kim, S.P. Generalized Vaidya spacetime in Lovelock gravity and thermodynamics on the apparent horizon. \textit{Phys. Rev. D} \textbf{2008}, \textit{78}, 124012. 

\bibitem{[39]} 
Rudra, P.; Biswas, R.; Debnath, U. Gravitational collapse in generalized Vaidya spacetime for Lovelock gravity theory. \textit{Astrophys. Space Sci.} \textbf{2011}, \textit{335}, 505-513. 

\bibitem{[40]} 
Dadhich, N.K.; Ghosh, S.G.; Jhingan, S. Gravitational collapse in pure Lovelock gravity in higher dimensions. \textit{Phys. Rev. D} \textbf{2013}, \textit{88}, 084024. 

\bibitem{byron} 
Maharaj, S.D.; Brassel, B.P. Junction conditions for composite matter in higher dimensions. \textit{Class. Quantum Grav.} \textbf{2021}, \textit{38}, 195006. 

\bibitem{suss1} 
Sussman, R.A. Weighed scalar averaging in LTB dust models: part I. Statistical fluctuations and gravitational entropy. \textit{Class. Quantum Grav.} \textbf{2013}, \textit{30}, 065015.

\bibitem{suss2} 
Sussman, R.A.; Larena, J. Gravitational entropies in LTB dust models. \textit{Class. Quantum Grav.} \textbf{2014}, \textit{31}, 075021. 

\bibitem{suss3} 
Sussman, R.A.; Larena, J. Gravitational entropy of local cosmic voids. \textit{Class. Quantum Grav.} \textbf{2015}, \textit{32}, 165012. 

\bibitem{HB} 
Hosoya, A.; Buchert, T.; Morita, M. Information entropy in cosmology. \textit{Phys. Rev. Lett.} \textbf{2004}, \textit{92}, 141302. 

\bibitem{LNP} 
Lima, R.C.; Nogales, J.A.C.; Pereira, S.H. Gravitational entropy of wormholes with exotic matter and in galactic halos. \textit{Int. J. Mod. Phys. D} \textbf{2020}, \textit{29}, 2050015. 

\bibitem{Piza} 
Pizaña, F.A.; Sussman, R.A.; Hidalgo, J.C. Gravitational entropy in Szekeres class I models. \textit{Class. Quantum Grav.} \textbf{2022}, \textit{39}, 185005. 

\bibitem{olivera1} 
de Oliveira, A.K.G.; Santos, N.O.; Kolassis, C.A. Collapse of a radiating star. \textit{Mon. Not. R. Astron. Soc.} \textbf{1985}, \textit{216}, 1001-1011. 

\bibitem{olivera2} 
de Oliveira, A.K.G.; Santos, N.O. Nonadiabatic gravitational collapse. \textit{Astrophys. J.} \textbf{1987}, \textit{312}, 640-645. 

\bibitem{[42]} 
Mansouri, R.; Nayeri, A. Gravitational coupling constant in arbitrary dimension. \textit{Gravit. Cosmol.} \textbf{1998}, \textit{4}, 142-144.

\bibitem{[43]} 
Sheykhi, A.; Moradpour, H.; Riazi, N. Lovelock gravity from entropic force. \textit{Gen. Relativ. Gravit.} \textbf{2013}, \textit{45}, 1033-1049. 

\bibitem{suss0} 
Sussman, R.A. New solutions for heat conducting fluids with a normal shear-free flow. \textit{Class. Quantum Grav.} \textbf{1993}, \textit{10}, 2675.

\bibitem{X1} 
Paliathanasis, A.; Govender, M.; Leon, G. Temporal evolution of a radiating star via Lie symmetries. \textit{Eur. Phys. J. C} \textbf{2021}, \textit{81}, 718. 

\bibitem{X2} 
Ivanov, B.V. Linear and Riccati equations in generating functions for stellar models in general relativity. \textit{Eur. Phys. J. Plus} \textbf{2020}, \textit{135}, 377. 

\bibitem{ray} 
Raychaudhuri, A.K.; Maiti, S.R. Conformal flatness and the Schwarzschild interior solution. \textit{J. Math. Phys.} \textbf{1979}, \textit{20}, 245-246. 

\bibitem{gold1} 
Glass, E.N.; Goldman, S.P. Relativistic spherical stars reformulated. \textit{J. Math. Phys.} \textbf{1978}, \textit{19}, 856-859. 

\bibitem{gold2} 
Goldman, S.P. Physical solutions to general relativistic fluid spheres. \textit{Astrophys. J.} \textbf{1978}, \textit{226}, 1079-1086. 

\bibitem{engycon}
Brassel, B.P.; Maharaj,  S.D.; Goswami, R. Higher-dimensional inhomogeneous composite fluids: energy conditions. \textit{Prog. Theor. Exp. Phys.} \textbf{2021}, 2021, 103E01.

\bibitem{[49]} 
Misner, C.W.; Sharp, D.H. Relativistic equations for adiabatic, spherically symmetric gravitational collapse. \textit{Phys. Rev.} \textbf{1964}, \textit{136}, B571-B576.

\bibitem{[50]} 
Cahill, M.E.; Taub, A.H. Spherically symmetric similarity solutions of the Einstein field equations for a perfect fluid. \textit{Commun. Math. Phys.} \textbf{1971}, \textit{21}, 1-40. 

\bibitem{[44]} 
Herrera, L.; Santos, N.O. Local anisotropy in self-gravitating systems. \textit{Phys. Rep.} \textbf{1997}, \textit{286}, 53-130. 

\bibitem{[45]} 
Israel, W. Singular hypersurfaces and thin shells in general relativity. \textit{Nuovo Cim. B} \textbf{1966}, \textit{44}, 1-14. 

\bibitem{[46]} 
O'Brien, S.; Synge, J.L. Jump conditions at discontinuities in general relativity. \textit{Commun. Dublin Inst. Adv. Stud.} \textbf{1952}, \textit{A9}, 1-20.

\bibitem{[47]} 
Lichnerowicz, A. \textit{Théories relativistes de la gravitation et de l'électromagnétisme}; Masson: Paris, 1955.

\bibitem{naked}Joshi, P.S.; Dadhich, N.; Maartens, R. Why do naked singularities form in gravitational collapse? \textit{Phys. Rev. D} \textbf{2002}, 65, 101501.
\bibitem{shear}  Joshi, P.S.; Goswami, R.; Dadhich, N. The critical role of shear in gravitational collapse. \textit{arXiv:gr-qc/} \textbf{2004} 0308012 [gr-qc].
\bibitem{ndimcollap} Goswami, R.; Joshi, P.S. Spherical gravitational collapse in $N$ dimensions. \textit{Phys. Rev. D} \textbf{2007}, 76, 084026.

\bibitem{split} 
Herrera, L.; Ospino, J.; Di Prisco, A.; Fuenmayor, E.; Troconis, O. Structure and evolution of self-gravitating objects and the orthogonal splitting of the Riemann tensor. \textit{Phys. Rev. D} \textbf{2009}, \textit{79}, 064025. 

\bibitem{[52]} 
Ivanov, B.V. A different approach to anisotropic spherical collapse with shear and heat radiation. \textit{Int. J. Mod. Phys. D} \textbf{2016}, \textit{25}, 1650049. 

\bibitem{[53]} 
Maharaj, S.D.; Tiwari, A.K.; Mohanlal, R.; Narain, R. Riccati equations for bounded radiating systems. \textit{J. Math. Phys.} \textbf{2016}, \textit{57}, 092501. 

\bibitem{[54]} 
Maartens, R. Dissipative cosmology. \textit{Class. Quantum Grav.} \textbf{1995}, \textit{12}, 1455.

\bibitem{[55]} 
Govender, M.; Maartens, R.; Maharaj, S.D. Relaxational effects in radiating stellar collapse. \textit{Mon. Not. R. Astron. Soc.} \textbf{1999}, \textit{310}, 557-564. 

\bibitem{[56]} 
Govinder, K.S.; Govender, M. Causal solutions for radiating stellar collapse. \textit{Phys. Lett. A} \textbf{2001}, \textit{283}, 71-79. 

\bibitem{[57]} 
Naidu, N.F.; Govender, M.; Govinder, K.S. Thermal evolution of a radiating anisotropic star with shear. \textit{Int. J. Mod. Phys. D} \textbf{2006}, \textit{15}, 1053-1065. 

\bibitem{[58]} 
Thirukkanesh, S.; Moopanar, S.; Govender, M. The final outcome of dissipative collapse in the presence of $\Lambda$. \textit{Pramana} \textbf{2012}, \textit{79}, 223-232. 

\end{thebibliography}
\end{document}